\documentclass[aps,prl,twocolumn,10pt,amsmath,superscriptaddress,amssymb,longbibliography]{revtex4}
\usepackage[normalem]{ulem}
\usepackage{epsfig}
\usepackage{epstopdf}
\usepackage{dcolumn}
\usepackage{amsbsy}
\usepackage{bm}
\usepackage{color,CJK}
\usepackage{amsmath,amssymb,amsfonts}
\usepackage{appendix}
\usepackage{graphicx}
\usepackage{algorithm}
\usepackage{enumerate}
\usepackage{makeidx}
\usepackage{braket}
\usepackage[usenames,dvipsnames]{xcolor}
\usepackage[driverfallback=dvipdfmx,colorlinks,linkcolor=blue,citecolor=blue,urlcolor=blue]{hyperref}

\usepackage{pifont}
\usepackage{graphicx}
\usepackage{textcomp}
\usepackage{helvet}

\begin{document}

\begin{CJK*}{GB}{gbsn}

\title{Observation of quantized vortex in atomic Bose-Einstein condensate at Dirac point with emergent spin-orbit coupling}
\author{Yunda Li}
\email[These authors contributed equally to this work.]{}
\author{Wei Han}
\email[These authors contributed equally to this work.]{}
\author{Zengming Meng}
\email[These authors contributed equally to this work.]{}
\author{Wenxin Yang}
\affiliation{State Key Laboratory of Quantum Optics and Quantum
Optics Devices, \\  Institute of Opto-Electronics, \\ Collaborative Innovation Center of Extreme Optics, Shanxi
University, Taiyuan, Shanxi 030006, People's Republic of China }
\author{Cheng Chin}
\email[Corresponding author email: ]{cchin@uchicago.edu}
\affiliation{James Franck Institute, Enrico Fermi Institute, Department of Physics,
University of Chicago, Chicago, Illinois 60637, USA}
\author{Jing Zhang}
\email[Corresponding author email: ]{jzhang74@sxu.edu.cn}
\affiliation{State Key Laboratory of Quantum Optics and Quantum
Optics Devices, \\  Institute of Opto-Electronics, \\ Collaborative Innovation Center of Extreme Optics, Shanxi
University, Taiyuan, Shanxi 030006, People's Republic of China }
\affiliation{Hefei National Laboratory, Hefei, Anhui 230088, People's Republic of China}

\maketitle

\end{CJK*}

\textbf{ When two or more energy bands become degenerate at a singular point in the momentum space, such singularity, or ``Dirac points", gives rise to intriguing quantum phenomena as well as unusual material properties. Systems at the Dirac points can possess topological charges and their unique properties can be probed by various methods, such as transport measurement, interferometry and momentum spectroscopy. While the topology of Dirac point in the momentum space is well studied theoretically, observation of topological defects in a many-body quantum systems at Dirac point remain an elusive goal. Based on atomic Bose-Einstein condensate in a graphene-like optical honeycomb lattice, we directly observe emergence of quantized vortices at the Dirac point. The phase diagram of lattice bosons at the Dirac point is revealed. Our work provides a new way of generating vortices in a quantum gas, and the method is generic and can be applied to different types of optical lattices with topological singularity, especially topological flat band near Dirac point for twisted bilayer optical lattices.}

Materials with gapless excitations such as semimetals~\cite{RevModPhys.93.025002,Ong2021} have attracted a great deal of interest in the past decade. A prominent example is graphene ~\cite{Geim2007,RevModPhys.81.109}, where a honeycomb lattice of carbon atoms supports singular Dirac points near which the energy dispersion is linear. The presence of singularity in momentum space leads to remarkable phenomena, such as trembling motion (Zitterbewegung) \cite{PhysRevLett.94.206801,PhysRevLett.100.113903}, Klein paradox \cite{Katsnelson2006}, electron chirality and anomalous quantum Hall effect \cite{Zhang2005}. The extraordinary properties of graphene near the Dirac point have motivated great interest to quantum simulate graphene based on electrons, photons and cold atoms~\cite{Polini2013}.

\begin{figure*}
\includegraphics[width=6.7in]{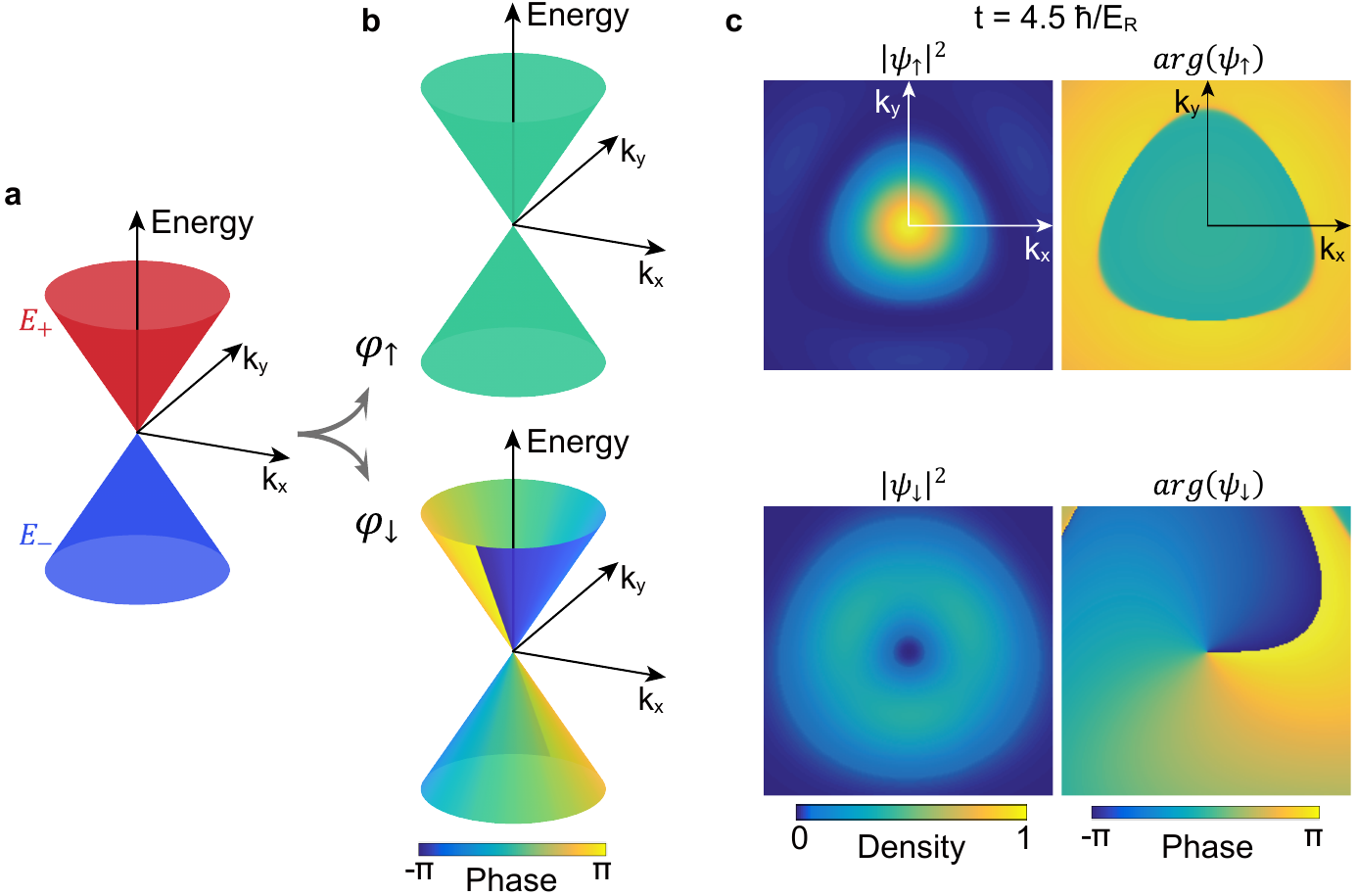}
\caption{ \textbf{Quantized vortex state at Dirac point.}~$\bf{a}$~Energy band near the Dirac point. Two eigenenergies of the Dirac point Hamiltonian $H_{D}=(k_{y}\sigma_{x}-k_{x}\sigma_{y})\nu_{0}= \Omega\sigma_{+}+\Omega^{*}\sigma_{-}$ are $E_{\pm}=\pm\nu_{0}|k|$ (upper and lower branches labeled with red and blue color, respectively). Here the two energy bands are assumed to touch at $k=0$. $\bf{b}$~Phase information of eigenvectors $\varphi_{\pm}=(\varphi_{\uparrow}, \varphi_{\downarrow})^{T}_{\pm}=(1,\mp ie^{i\theta})^{T}$ of the Dirac point Hamiltonian associated with energies $E_{\pm}$ are plotted. $\bf{c}$~Generation of a quantized vortex state with topological charge $l=-1$ at Dirac point. A BEC with about 0.2 $k_{R}$ momentum width initially prepared in the spin up state is quenched to the Dirac point of the honeycomb lattice with lattice depth $V_0=12~E_R$. The dynamical evolution generates amplitude in the spin down component that carries a quantized vortex with topological charge $l=-1$. The atomic density distribution (amplitude) and phases of two spin components of the state $ (\psi_{\uparrow}, \psi_{\downarrow})^{T}$ are shown at time $t=4.5\ \hbar/E_R$ after the quench. The broken rotational symmetry of the density resulting from the effect of additional high-order term of the effective Hamiltonian (see the supplement in detail).}
\label{Fig1}
\end{figure*}

Quantum physics at the Dirac points has been investigated in photonic crystal systems. Dirac points in photonic graphene offer a unique way to generate optical vortices (optical beam carrying orbital angular momenta) by transforming pseudo-spin winding in momentum space into phase winding of the optical beam in real space~\cite{PhysRevLett.98.103901,Song_2015,Song2015-1}.

In quantum gases, the optical honeycomb lattice has also been realized experimentally~\cite{Soltan-Panahi2011,Tarruell2012}, and the Dirac point and global topology can be determined by the evolution of quantum states via transporting wave packets~\cite{Duca-2015,Li2016,Brown2022}, and momentum-resolved tomography~\cite{Flaschner2016,Weinberg_2016,PhysRevLett.118.240403,Flaschner2018,Tarnowski2019}. In these systems Dirac point can be identified from the non-Abelian transformation when the system passes through the singularity~\cite{Brown2022}. However, formation of quantized vortices or topological defects in a quantum gas at the Dirac point has not yet been directly observed.

\begin{figure*}
\includegraphics[width=6.7in]{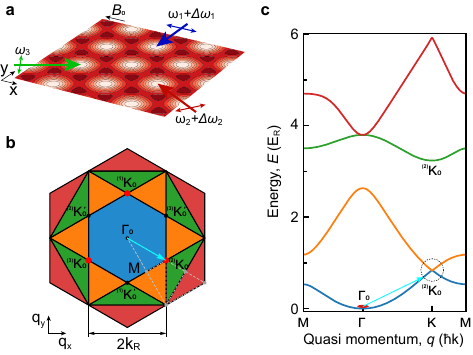}
\caption{ \textbf{Honeycomb lattice setup and energy band structure.}~$\bf{a}$~We create the honeycomb lattice by interfering three laser beams at an angle of $120^{\circ}$ with in-plane polarization. A homogeneous magnetic field $B_{0}$ is applied in the $x$ direction. The thick arrows indicate the propagating direction of the light in the x-y plane, and the thin arrows indicate the linear polarization of the light. $\bf{b}$~The first four Brillouin zones of the lattice are presented with blue, orange, green, red colors, respectively. Dirac points in the first Brillouin zone are labeled as $^{(a)}K_{0}$ or $^{(a)}K'_{0}$, where $a=1,2,3$. The dashed lines show the high symmetry paths for the band structure of $\bf{c}$. $\bf{c}$ The energy bands of the honeycomb lattice with the lattice depth 12~$E_R$. The colors of energy bands correspond to those in panel $\bf{b}$. A Dirac point occurs at $K_{0}$ (dashed circle), where the lowest two bands touch linearly, and can be described by the model presented in Fig.~1. In our experiment, the condensate (red ellipse) is initially prepared in the lowest state $\Gamma_0$ and then quenched to the Dirac point at $^{(2)}K_0$, indicated by the cyan arrow.}
\label{Fig1}
\end{figure*}

In this work, we observe quantized vortex generation in a Bose-Einstein condensate (BEC) of Rubidium-87 ($^{87}$Rb) atoms prepared at the Dirac point of a two-dimensional graphene-like honeycomb lattice. The Hamiltonian of the system can be expressed in the Rashba spin-orbit coupling form with $H_{R}=(k_{y}\sigma_{x}-k_{x}\sigma_{y})\nu_{0}\equiv \Omega\sigma_{+}+\Omega^{*}\sigma_{-}$, where $(k_{x},k_{y})=(k \cos \theta, k \sin \theta) $ is the momentum operator, $\sigma_{x,y}$ are Pauli matrices, $\Omega=i k e^{-i\theta}\nu_{0}$, $\nu_{0}=\hbar^{2}k_{0}/2m$ is the spin-orbit coupling strength, $\sigma_{\pm}=(\sigma_{x} \pm i\sigma_{y})/2$ are the spin raising/lowering operators, $m$ is the atomic mass, $\hbar$ is the reduced Planck constant, and $k_{0}=2\pi
/\lambda$ is the photon recoil momentum of the lattice beam at wavelength $\lambda$. This Hamiltonian can generate a quantized vortex at the Dirac point. The eigenstates and their spin projections are shown in Fig.~1.

We create the optical honeycomb lattice by interfering three red-detuned laser beams operating at wavelength $\lambda=820$~nm. The beams propagate within the x-y plane and intersect at an angle $\xi=120^{\circ}$ with linear polarizations in the x-y plane, as shown in Fig.~2a. A homogeneous magnetic bias field $B_{0}=2.7$~G is applied in the $x$ axis, which defines the quantization axis. The interference of the laser beams leads to an attractive scalar lattice potential on the atoms~\cite{Weinberg_2016,Wen:21}. A honeycomb lattice forms with in-plane polarizations of the lattice beams. The natural momentum and energy units of the lattice are the recoil momentum $\hbar k_{R}$ and the recoil energy $E_{R}=\hbar^2 k_{R}^2/2m = h\times 2.56$~kHz, where $k_{R}=k_{0}\sin(\xi/2)$ is the lattice momentum. The lowest $s$- and $p$- energy bands of the honeycomb lattice exhibit Dirac points with linear dispersion, shown in Figs.~2b and 2c.

The atoms are initially prepared in the spin state $\left\vert F=1, m_F=1\right\rangle$. The sample is confined in a single layer of an accordion lattice in the $z$-direction (gravity direction), and then loaded into the lowest Bloch band at the $\Gamma_{0}$-point by adiabatically ramping up the honeycomb lattice. At the same time, a harmonic trapping potential is added in the $x$-$y$ direction. Next, we smoothly accelerate the honeycomb lattice potential toward the velocity $\mathbf{v}$. The low acceleration ensures that the atoms remain in the $s$-orbital band manifold with zero quasi-momentum in the laboratory frame while the atoms evolve adiabatically into the quasi-momentum $\mathbf{q}=-m \mathbf{v}/\hbar$ in the lattice frame. Here we control the lattice velocity $\mathbf{v}_{i}$ by detuning the $i$-th lattice beam according to $\mathbf{v}_{i}=a\Delta \omega_{i}\hat{k}_{i}$, where $a=2\lambda/3$ is lattice constant and $\hat{k}_i$ is the unit vector in the propagating direction of the $i$-th laser beam that forms the honeycomb lattice.

\begin{figure*}
\includegraphics[width=6in]{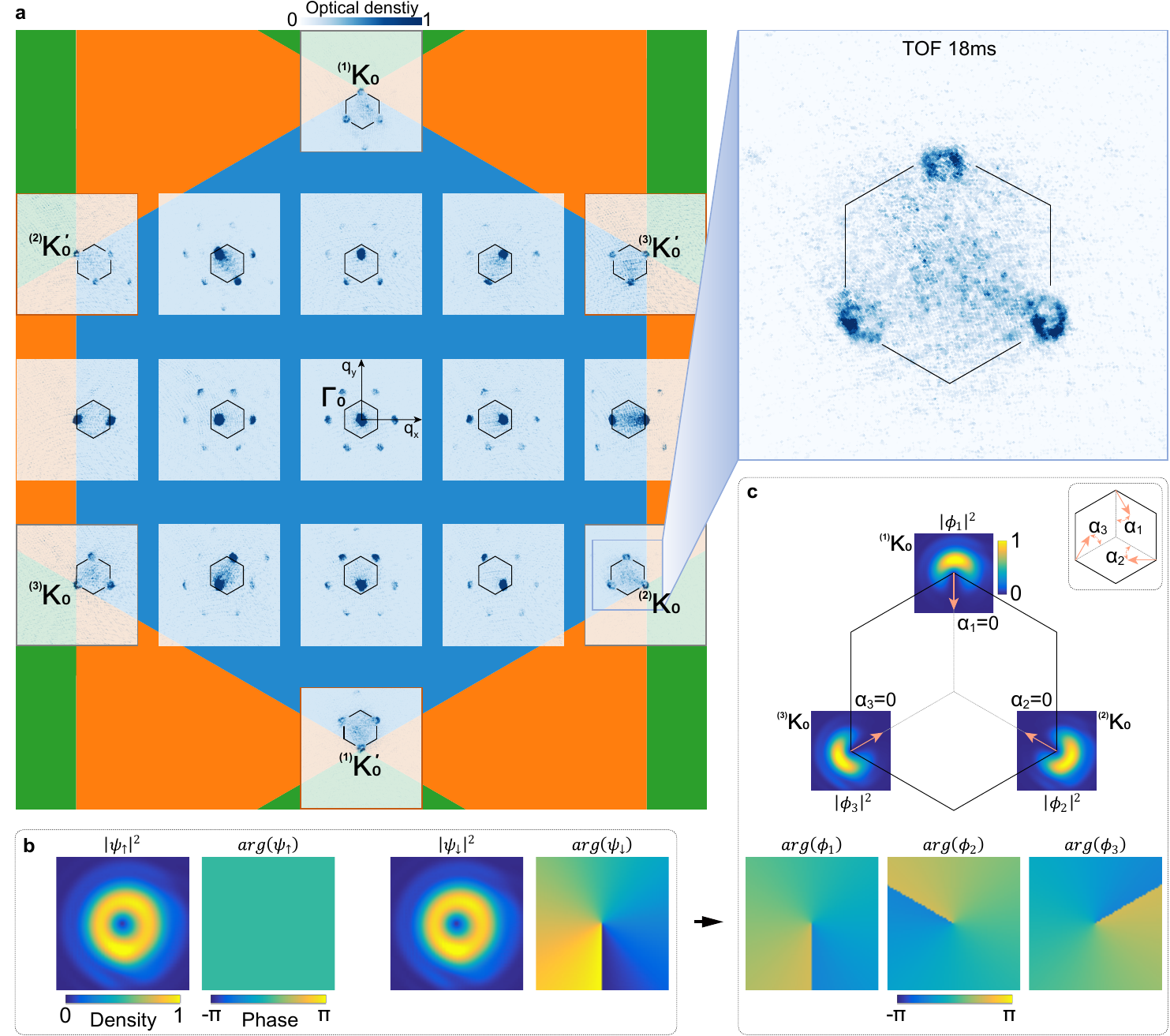}
\caption{ \textbf{Observation of vortex in a Bose-Einstein condensate at Dirac point.} $\bf{a}$~Time-of-flight images of BECs at different quasi-momentum states. Initially prepared in the ground state $\Gamma_{0}$, the BECs are adiabatically transferred to various states, including the six Dirac points at the corners of the first Brillioun zone where vortex emerges. The filling colors of the Brillouin zones are consistent with those in Fig.~2b. The centres of the TOF image panels are pinned at the positions of Brillouin zone where the BECs are prepared.
The small hexagon in each panel labels the first Brillouin zone boundary in the atomic momentum distribution. The magnified image details the vortex structure of the BECs with the topological charge $l=-1$. Here BEC is projected to three momentum components $^{(1)}K_{0}$, $^{(2)}K_{0}$ and $^{(3)}K_{0}$ during the TOF expansion. The image shows a directional opening $\alpha_{a}=0$ ($a=1,2,3$), which comes from the interference between two pseudo-spin components $\mathbf{\psi}_{\uparrow}$ and $\mathbf{\psi}_{\downarrow}$ of the generated state, see Supplementary Material. The depth of optical lattice is 12~$E_{R}$ and the harmonic trapping frequency in the $x$-$y$ direction is $2\pi\times40$~Hz. $\bf{b}$~Calculated atomic density and phase of the pseudo-spin components $\psi_{\uparrow}$ and $\psi_{\downarrow}$ of the generated state for a BEC transferred to the Dirac point $K_{0}$. $\bf{c}$ Calculated three projections of the BEC at Dirac point $\phi_a(k)=\langle^{(a)}K_0+k|\psi\rangle$ relative to the momenta $^{(a)}K_0$. The projections describe the magnified TOF image of (a) and are connected to the quasi-spin components of the generated state as $\phi_{1}=\frac{1}{\sqrt{3}}\left( \psi _{\uparrow}+\psi _{\downarrow}\right)$, $\phi_{2}=\frac{1}{\sqrt{3}}\left( e^{i\frac{2\pi }{3}}\psi _{\uparrow}+e^{-i\frac{2\pi}{3}}\psi _{\downarrow}\right)$, and $\phi_{3}=\frac{1}{\sqrt{3}}\left( e^{-i\frac{2\pi}{3}}\psi _{\uparrow}+e^{i\frac{2\pi }{3}}\psi _{\downarrow}\right)$.}
\label{Fig3}
\end{figure*}

In order to create quantized vortex at the Dirac point, we transfer the BEC adiabatically from the ground state $\Gamma_{0}$ to Dirac point $K_{0}$. Since the Dirac point is not the ground state of the system, the BECs have a limited lifetime of several milliseconds (see Supplementary Material). The lifetime is, however, sufficient since the BEC can be adiabatically transferred from $\Gamma_{0}$ to $K_{0}$ within 0.8~ms. At the same time, we introduce a harmonic trapping potential in x-y direction, and then map the momentum distribution of the BEC based on time-of-flight (TOF) imaging. Here the harmonic trapping potential is critical for the formation of vortex, as it does not commute with the the pseudo-spin-orbit coupling and thus induces a gauge potential at the Dirac point (see Supplemental Material). We map out the entire first Brillouin zone $n=1$, including the six Dirac points $^{(a)}K_{0}$ and $^{(a)}K'_{0}$, ($a=1,2,3$), see Fig.~3a, which connect to the second energy bands $n=2$ as shown in Fig.~2c. In TOF images, the vortex structure manifests in the density distribution of the three distinct momentum components when the BEC is located at the Dirac points of the first Brillouin zone, see the enlarged image of Fig.~3a.

To model the vortex formation at the Dirac point, we introduce an effective Hamiltonian with Rashba pseudo-spin-orbit coupling (see details in Supplementary Material). To describe the BEC near the Dirac points $K_{0}$ ($K'_{0}$) of the lowest band we set the frequency detunings of the lattice beams as $\Delta\omega_{2}=\pm4E_{R}/h$ ($+$ for $K_{0}$ and $-$ for $K'_{0}$). This condition corresponds to the two-photon Bragg transition frequency $\omega_{B}$ between lattice beams 2 and 1 (3). The quadratic energy-momentum dispersion of the massive atoms defines a unique energy difference $\omega_{B}=4E_{R}/\hbar$, which just matches the stimulated two-photon Bragg transition frequency. In this case, the three momentum components of the atoms at $^{(1)}K_{0}$, $^{(2)}K_{0}$ and $^{(3)}K_{0}$ are resonantly coupled, and the effective Hamiltonian shows three modes with pair-wise couplings, which in the dressed state basis is given by

\begin{eqnarray}
H_{eff}&=&\frac{\hbar^{2}k_{0}}{2m}\left(
\begin{array}{ccc}
0 & k_{y}+ik_{x} & k_{y}-ik_{x} \\
k_{y}-ik_{x} & 0 & k_{y}+ik_{x} \\
k_{y}+ik_{x} & k_{y}-ik_{x} & 0%
\end{array}%
\right)
 \nonumber \\
&&+\left(
\begin{array}{ccc}
-\Lambda & 0 & 0 \\
0 & -\Lambda & 0 \\
0 & 0 & 2\Lambda%
\end{array}%
\right)
,\label{SU3Hamiltonian}
\end{eqnarray}
where $\Lambda$ is the pair-wise coupling strength and proportional to the lattice depth. This Hamiltonian resembles the ring scheme of three cyclically Raman coupled ground hyperfine
spin states~\cite{PhysRevA.81.053403,PhysRevA.84.025602,Huang2016a}. The dressed-state basis $\hat{\Psi}(\mathbf{k})$ is connected to the bare-state basis $\hat{\Phi}(\mathbf{k})=\{|^{(1)}K_{0}+\mathbf{k}\rangle,|^{(2)}K_{0}+\mathbf{k}\rangle,|^{(3)}K_{0}+\mathbf{k}\rangle\}$ through a unitary transformation
\begin{equation}
S=\frac{1}{\sqrt{3}}\left(
\begin{array}{ccc}
1 & 1 & 1 \\
e^{i\frac{2\pi }{3}} & e^{-i\frac{2\pi }{3}} & 1 \\
e^{-i\frac{2\pi }{3}} & e^{i\frac{2\pi }{3}} & 1%
\end{array}%
\right).
\end{equation}

\begin{figure*}
\includegraphics[width=6.5in]{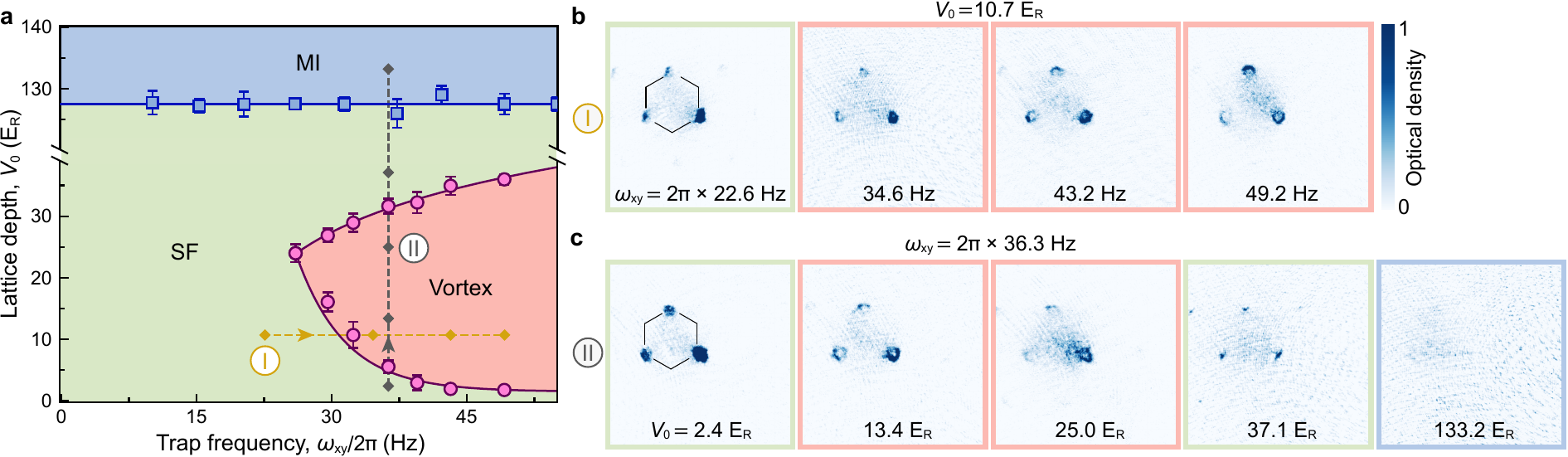}
\caption{ \textbf{Quantum phases of Bose gases at the Dirac point of a honeycomb optical lattice.} $\bf{a}$~Phase diagram for different trap frequencies $\omega_{xy}$ and lattice depth $V_0$. Here SF and MI refer to superfluid (green) and Mott insulator (blue), respectively. The red area is the region where vortices form in the SF. Error bars show the standard deviations of the measured phase boundary position. Solid lines are guides to the eye. For comparison, theoretical calculation of the phase diagram is shown in Fig.~S6, see Supplementary Material. Examples of images along the line
{\fontfamily{phv}\selectfont\raisebox{0.2ex}{\scalebox{0.95}{\textcircled{\raisebox{-0.2ex}{\scalebox{0.85}{I}}}}}}
with fixed lattice depth of $V_0=$ 10.7 $E_{R}$ and increasing harmonic trapping frequencies in the $x$-$y$ are shown in panel~$\bf{b}$. Images along the line
{\fontfamily{phv}\selectfont\raisebox{0.2ex}{\scalebox{0.95}{\textcircled{\raisebox{-0.2ex}{\scalebox{0.85}{II}}}}}}
with fixed harmonic trapping frequency $\omega_{xy}=2\pi\times36.3$ Hz and increasing lattice depths are shown in panel~$\bf{c}$. Color frames of the images from both panels $\bf{b}$ and $\bf{c}$ show assignment of the quantum phase according to the color code in panel~ $\bf{a}$.}
\label{Fig4}
\end{figure*}

This effective Hamiltonian very well describes the $n=1, 2, 3$ Bloch bands near $K_{0}$ and the singular Dirac degeneracy between the $n=1$ and $n=2$ bands as shown in Fig.~2c. Full numerical calculation of the energy bands near the $K_{0}$ is shown in Fig. S5 of Supplementary Material. The highest energy band has a much higher energy 3$\Lambda$ than the lower two bands in Eq.~(1). In an adiabatic process, we can neglect amplitude in the highest energy band and consider only atoms in the lowest two dressed states near the degeneracy. Thus a pseudo-spin 1/2 Hamiltonian with the Rashba pseudo-spin-orbit coupling becomes a very good approximation near the Dirac point and yields the linear dispersion. On the other hand, an effective Hamiltonian with the Dresselhaus spin-orbit coupling with opposite sign of topological charge describes the linear Dirac equation near $K'_{0}$, see Supplementary Material for details.

The conventional method to create vortices near the Dirac point is based on dynamical process of coherent spin flip, see Fig.~1c (or Fig.~S8 in the Supplementary Material for details). Here the BEC with finite momentum is initially prepared in the spin up (or spin down) component in the pseudo-spin 1/2 representation at the Dirac point $K_{0}$. The spin is then flipped in the evolution, which generates spin down (or spin up) component that carries the quantized vortex with topological charge $l=-1$ (or $l=+1$), see Supplementary Material. This method has been employed to generate optical vortices in photonic graphene~\cite{Song_2015,Song2015-1}.

Here we develop a new method to create quantized vortex at Dirac point by a gauge potential induced by the noncommutativity between the pseudo spin-orbit coupling and the harmonic trapping potential in the $x$-$y$ direction. This method requires adiabatic evolution of the atoms, which is unique for ultracold atom system and provides the controllable parameters to observe quantum phase transitions. The induced gauge potential is momentum-dependent, and yields a angular superfluid velocity and vorticity. When the BEC is accelerated adiabatically from $\Gamma_{0}$ to $K_{0}$ ($K'_{0}$), with an appropriate harmonic trapping frequency and optical lattice depth, the condensate is prepared at the spin eigenstate $\Psi_{s}=(1,-i e^{\pm i\theta})^{T}$ in the pseudo-spin 1/2 representation ($+\theta$ for $K_{0}$ and $-\theta$ for $K'_{0}$), which is composed of a phase trivial spin up component and a phase nontrivial spin down component. In this case, a quantum vortex with an orbital angular momenta of $\pm\hbar$ creates naturally as indicated in Fig.~1b (see Supplementary Material for details).

We observe the vortex of BECs at the Dirac point based on TOF images. The quantized vortex is observed in three momentum components with triangular configuration locating at the Dirac points of the first Brillouin zone as shown in the magnified image of Fig.~3a. We can see the quantized vortex is off-center in the bare state representation due to the interference between two spin dressed states. Note that one of two spin dressed states has the phase singularities $e^{i \theta}$ for $K_{0}$ and the other is trivial, see Fig.~3b. The off-center atomic density distribution in the bare-state basis for $K_{0}$ acquires a directional opening with an angle $\alpha=0$, see Fig.~3c, where the angle $\alpha$ is defined relative to the axis pointing to the center of the hexagonal band. The experimental result is in excellent agreement with our theoretical calculation. Although $K_{0}$ and $K'_{0}$ have opposite signs of the topological charges, the atomic density distribution in the bare-state basis for $K'_{0}$ has the same directional opening with $\alpha=0$ but different momentum components.

Our scheme also offers a new and convenient method to probe the characteristics and singularities in the band structure. We can transfer BECs from $\Gamma_{0}$ to an any arbitrary quasi-momentum state and obtain the information of the momentum components of the high-symmetry $K$ and M. For example, M point is characterized by two momentum components. Especially, we have identified the singularity of all the six Dirac points $^{(a)}K_{0}$ and $^{(a)}K'_{0}$ with $a=1,2,3$ at the corners of the first Brillioun zone by the emergence of quantum vortex. Three distinct momentum components with triangular configuration for $K_{0}$ are located at three Dirac points of the first Brillouin zone in the lattice frame, and those for $K'_{0}$ are located at another three Dirac points, as shown in Fig.~3a.

Near the Dirac point we also observe quantum phase transition based on two controllable parameters. Our system includes a pair of the non-commutative interactions:  harmonic potential and the the pseudo-spin-orbit coupling. Moreover this system is also described by the Bose-Hubbard model, which permits a quantum phase transition from superfluid (SF) to Mott insulator (MI) as the depth of optical lattices increases. Thus the two controllable parameters: optical lattice depth and the harmonic trapping potential in the $x$-$y$ direction can induce quantum phase transitions. When we maintain the time sequence of ramping the optical honeycomb lattice and the harmonic trapping laser in $x$-$y$ direction, three distinct quantum phases are expected, including regular SF, SF with a vortex, and MI, see Fig.~4a.

We observe no quantized vortex in a weak harmonic trap or in a deep lattice. From the TOF images for the different harmonic trapping frequencies at the lattice depth of 10.7 $E_{R}$ (see Fig.~4b), and for the various lattice depths with fixed trapping frequency at $2\pi\times36.3$~Hz (see Fig.~4c), we show good agreement with numerical calculation to confirm the three phases of the system based on the three-mode model (see Fig.~S6 in Supplementary Material). We find that the hole of quantized vortex becomes clearer and its size is larger when increase harmonic trapping potential in $x$-$y$ direction, which is consistent with our theoretical model. For deeper lattice, the atoms enter the MI region with no superfluid component due to the strong interaction and so there is no quantized vortex of matter wave.

To summarize, our research presents a novel scheme to probe quantum many-body systems with gauge field based on ultracold atoms prepared near the Dirac point. Quantized vortices of BEC generated by the Dirac singular point in momentum space is observed. The uniqueness of this method is that it does not involve any chirality or a winding structure in real space and thus position alignment is not required. The scheme we develop to prepare a BEC at the singular point in momentum space opens the door to the intriguing interplay between topology, quantum degeneracy and interactions based on cold atoms. Applying our method to other lattice geometries will offer access to exotic topological objects, such as higher-order vortices and Dirac monopole at 3D Weyl point. In recent years, the twisted bilayer-graphene has attracted broad attention, which hosts topological flat bands near Dirac point and exhibits remarkable interacting phases including correlated insulators, Chern insulators, and superconductor~\cite{Cao2018-2,Yankowitz1059,Lu2019}. At the same time, the twisted-bilayer optical lattices based on ultracold atoms have been realized experimentally by our group~\cite{Meng2023}. This work takes a first step to study BEC and superfluid in the topological flat bands near Dirac point and the related rich interacting phases~\cite{PhysRevLett.127.170404,Lukin2023}.

\begin{acknowledgments}
This research is supported by Innovation Program for Quantum Science and Technology (Grant No. 2021ZD0302003), NSFC (Grant No. 12034011, U23A6004, 12322409, 12474252, 12474266, 12374245), National Key Research and Development Program of China (Grant No. 2022YFA1404101, 2021YFA1401700) and Tencent (Xplorer Prize). CC acknowledges support by the National Science Foundation (Grant No. PHY-2409612).
\end{acknowledgments}

\bibliography{references}
\bibliographystyle{naturemag}

\begin{widetext}

\section*{SUPPLEMENTARY MATERIAL}

\subsection*{Details of the experimental setup}

A BEC of typical $5\times10^5$ atoms in $|F=1,m_{F}=1\rangle$ state is prepared in the crossed optical dipole trap. We load the 3D shaped BEC into a single layer of the 2D pancakes at maximum separation and then compress the pancake adiabatically to reach a deep 2D regime by an accordion lattice, as shown in Fig. S1. The accordion lattice is formed by a 532~nm laser beam deflected by an acousto-optic deflector (AOD) and then split into two beams with variable spacing adjusted by the AOD. The two beams are focused onto the atoms with a 150~mm aspherical lens and interfere to form a standing wave in the vertical direction with variable separation. This separation can be varied from 12~$\mu m$ down to 3~$\mu m$. We optimize it at $2\pi\times1$~kHz to achieve superfluid of ultracold atoms. The 2D pancakes provide the strong confinement in z-direction, which can apply balance forces to cancel or compensate the gravity. We use a far-red detuned 1064 nm laser propagating along the z direction to provide a harmonic trapping potential in x-y direction.

\begin{figure*}[b]
\includegraphics[width=3in]{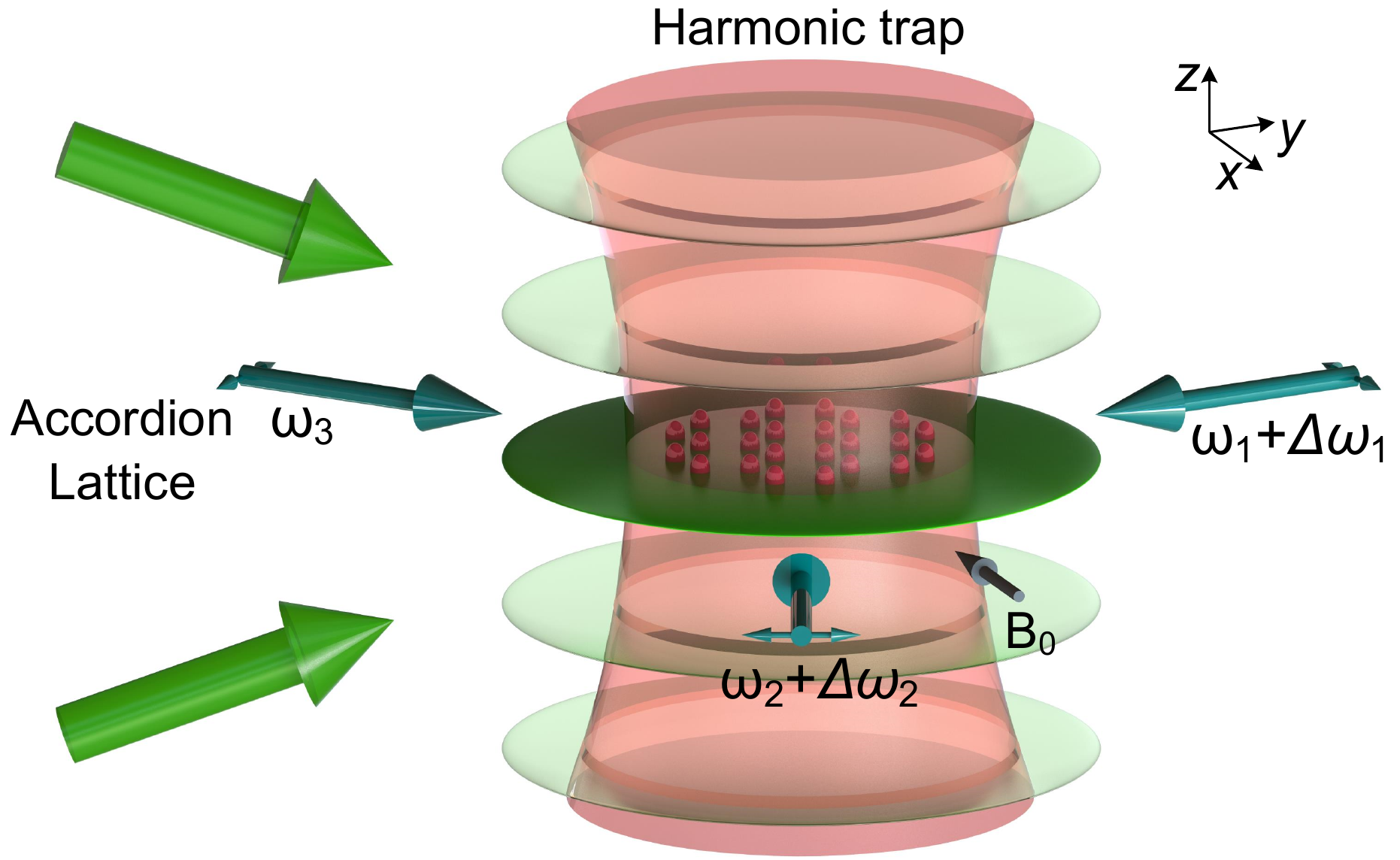}
\setcounter{figure}{0}
\renewcommand{\thefigure}{S\arabic{figure}}
\caption{\textbf{Schematic of experimental setup.} 3D shaped BEC is loaded into a single layer of the 2D pancakes by an accordion lattice. A far-red-detuned 1064 nm laser propagating along z direction provides a harmonic trapping potential in x-y direction. The optical honeycomb lattice is formed by the three red-detuned laser beams, which is controlled by the single-pass acousto-optic modulators. The beams propagate within the x-y plane and intersect at an angle $\xi=120^{\circ}$ with linear p- polarizations in the x-y plane. A homogeneous magnetic bias field is applied along the $x$ axis.}
\label{figS2}
\end{figure*}

The optical honeycomb lattice is formed by the three red-detuned laser beams operating at a wavelength $\lambda=820$ nm. Each lattice beam is controlled by a single-pass acousto-optic modulators (AOM), which is used to change the frequency detuning between lattice beams. Then the three lattice beams are coupled into
polarization-maintaining single-mode fibres respectively to improve the stability of the beam pointing and achieve better beam-profile quality. After
the fibres, three lattice beams intersect at relative angles of $120^{\circ}$ and interfere to create the optical honeycomb lattice potential. The lattice velocity $\mathbf{v}_{i}$ can be controlled by changing the detuning frequency of lattice beams $\mathbf{v}_{i}=a\Delta \omega_{i}\hat{k}_{i}$. Therefore, we can prepare BEC at the any quasi-momentum $\mathbf{q}=-m \mathbf{v}/\hbar$ in the lattice frame by choosing the appropriate linear combination of $\mathbf{v}_{1}$ and $\mathbf{v}_{2}$.

The ramp shapes of lattice beam intensity, detuning, harmonic trapping intensity in the x-y direction, and accordion lattice intensity are given in Fig. S2. After creating a BEC, we load the 3D shaped BEC into a single layer of the 2D pancakes at maximum separation and then compress the pancake adiabatically to reach a deep 2D regime by an accordion lattice. Then we load the atoms into the lowest Bloch band at the $\Gamma_{0}$ -point of the first Brillouin zone by adiabatically ramped up the honeycomb lattice. At the same time, the harmonic trapping potential is added in x-y direction. We transfer BEC adiabatically from $\Gamma_{0}$ to $K_{0}$, so the eigenstate $\Psi'_{s}=(1,0)^{T}$ in the gauge transformed representation is prepared adiabatically with the help of the harmonic trapping potential in x-y direction. Finally, the harmonic trapping potential is ramped off quickly within 0.2 ms, followed by simultaneous switching off of the honeycomb lattice and accordion lattice. The atoms are detected by absorptive imaging with TOF of 18 ms to map the momentum components into the spatial density distribution of the BEC.

\begin{figure*}[tb]
\includegraphics[width=3in]{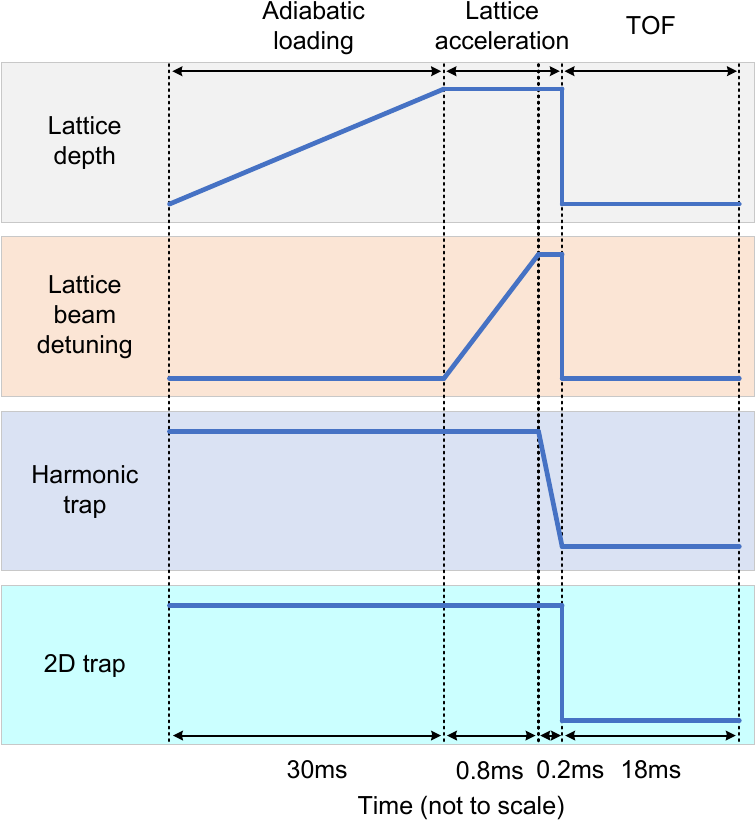}
\setcounter{figure}{1}
\renewcommand{\thefigure}{S\arabic{figure}}
\caption{\textbf{Experimental sequence.} Ramp shape of lattice beam intensity, detuning, harmonic trapping intensity in x-y direction, and accordion lattice intensity.}
\label{figS2}
\end{figure*}

\begin{figure*}[b]
\includegraphics[width=3in]{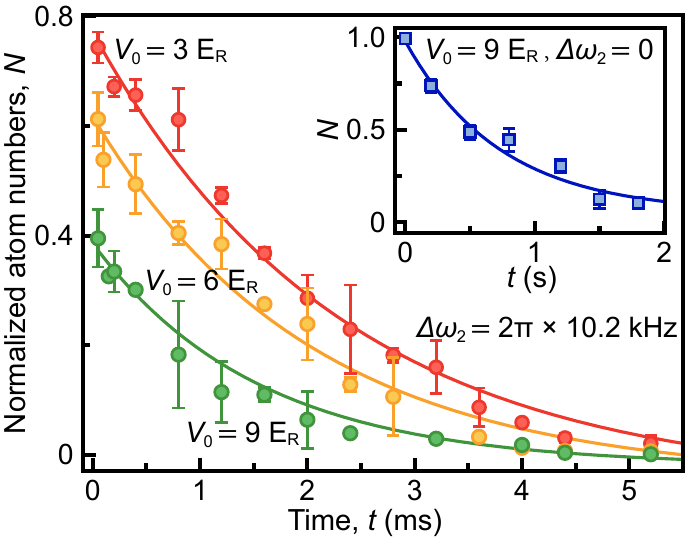}
\setcounter{figure}{2}
\renewcommand{\thefigure}{S\arabic{figure}}
\caption{\textbf{The lifetime of BEC at Dirac point.} The normalized atomic numbers of the superfluid component are plotted as function of waiting time at Dirac point for the different lattice depth. For comparison, the inset shows the data of the superfluid component as function of waiting time at the $\Gamma_{0}$ -point of the lowest Bloch band. Each point is based on three or more measurements and error bars show the standard deviations of the mean.}
\label{figS3}
\end{figure*}

We transfer BEC from $\Gamma_{0}$ to $K_{0}$ with 0.8 ms and measure the atomic number of superfluid at Dirac point as function of waiting time. BEC at Dirac point has the limited lifetime with about several milliseconds since this state is not real ground state of the system as shown in Fig. S3. For comparison, BEC at the $\Gamma_{0}$ -point of the lowest Bloch band has long lifetime with more than one second.

In the experiment, we measure the contrast from the momentum-space diffraction peaks in the TOF images to determine the phase boundary between SF and MI. The phase boundary between SF and vortex is identified by checking if there is a quantized vortex of matter wave in the TOF images. We count the atoms in the different momentum regions to calculate the contrast, which is defined as $(n_{max}-n_{min})/(n_{max}+n_{min})$~\cite{Gerbier2005}. The vortex can be visually identified in the TOF images. The experimental points of the phase boundaries is shown in Fig.~4a, which are measured by choosing the several paths similar to the line
{\fontfamily{phv}\selectfont\raisebox{0.2ex}{\scalebox{0.95}{\textcircled{\raisebox{-0.2ex}{\scalebox{0.85}{II}}}}}}
with fixed harmonic trapping frequency and increasing lattice depths. First, we roughly determine the phase boundary by changing the lattice depths with the step of 1 $E_{R}$. This procedure only performs one measurement to obtain a rough phase boundary. Then we determine the phase boundary precisely by changing the lattice depths with the step of 0.2 $E_{R}$ near the rough phase boundary. Each point is based on three or more measurements and error bars are obtained to show the standard deviations of the mean.

\subsection*{Theoretical model}

\noindent \textbf{1. The effective Hamiltonian at Dirac point}

The condensate dynamics in the honeycomb optical lattice can be described by the time-dependent Schr\"{o}dinger equation
\begin{equation}
i\hbar\partial_t|\chi\rangle=\mathcal{H}|\chi\rangle,\label{Schr1}
\end{equation}
where
\begin{equation}
\mathcal{H}=-\frac{\hbar^2}{2m}\mathbf{\nabla}^2+\frac{2}{9}V_0\big[\cos(\mathbf{G}_1\cdot \mathbf{r})+\cos(\mathbf{G}_2\cdot \mathbf{r})+\cos(\mathbf{G}_3\cdot \mathbf{r})\big],
\end{equation}
and $V_0$ is the lattice depth, $\mathbf{G}_1=-\sqrt{3}k_0 \hat{\mathbf{e}}_x$, $\mathbf{G}_2=\sqrt{3}k_0 \big(\frac{1}{2}\hat{\mathbf{e}}_x+\frac{\sqrt{3}}{2}\hat{\mathbf{e}}_y\big)$ and $\mathbf{G}_3=\sqrt{3}k_0 \big(\frac{1}{2}\hat{\mathbf{e}}_x-\frac{\sqrt{3}}{2}\hat{\mathbf{e}}_y\big)$ are the reciprocal lattice vectors from the first order two-photon transitions with $k_0=2\pi/\lambda$ being the wave number of the laser. Here, $m$ is the atomic mass and $\hbar$ is the Planck constant divided by $2\pi$.

The condensate with momentum $\mathbf{q}$ evolves according to the Bloch eigenstates $|\chi_{n,\mathbf{q}}\rangle$ of $\mathcal{H}$ as
\begin{equation}
|\chi_{\mathbf{q}}(t)\rangle=e^{-iE_nt/\hbar}|\chi_{n,\mathbf{q}}\rangle\langle\chi_{n,\mathbf{q}}|\chi_{\mathbf{q}}(0)\rangle.
\end{equation}
Expanding the Bloch state $|\chi_{n,\mathbf{q}}\rangle$ in the discrete plane wave basis
\begin{equation}
|\chi_{n,\mathbf{q}}\rangle=\sum_p a_\mathbf{q}(p)e^{i(\mathbf{q}+\mathbf{G}_p)\cdot \mathbf{r}},
\end{equation}
with $\mathbf{G}_p=k \mathbf{G}_1+l \mathbf{G}_2$ and $p={(k,l)}\in\mathbb{Z}^{2}$, and substituting it into the Schr\"{o}dinger equation, one can write the momentum-space Hamiltonian as
\begin{equation}
\mathcal{H}_{p',p}=\frac{\hbar^2}{2m}|\mathbf{q}+\mathbf{G}_p|^2\delta_{p',p}+\frac{2V_0}{3}\delta_{p',p}+\frac{V_0}{9}\delta_{p',p+\Delta p},\label{MomenHamiltonian}
\end{equation}
where $\Delta p\in{\{(1,0),(-1,0),(0,1),(0,-1),(1,1),(-1,-1)\}}$.

For the ground state of the condensate in a static optical lattice, most atoms reside in the zero quasi-momentum in the laboratory frame, and the atomic number decreases dramatically with the increase of diffraction order. This can be understood by the Raman detuning in the two-photon process for the diffraction, which results from the change of the atomic kinetic energy. For the first-order diffraction, the Raman detuning is $\frac{\hbar^2(2k_R)^2}{2m}=4E_R$, where $k_R=k_0\sin\frac{\xi}{2}$ with $\xi=\frac{2\pi}{3}$ being the angle between two lasers, and the detuning increases with the increase of the diffraction order.

\begin{figure*}[tb]
\includegraphics[width=4in]{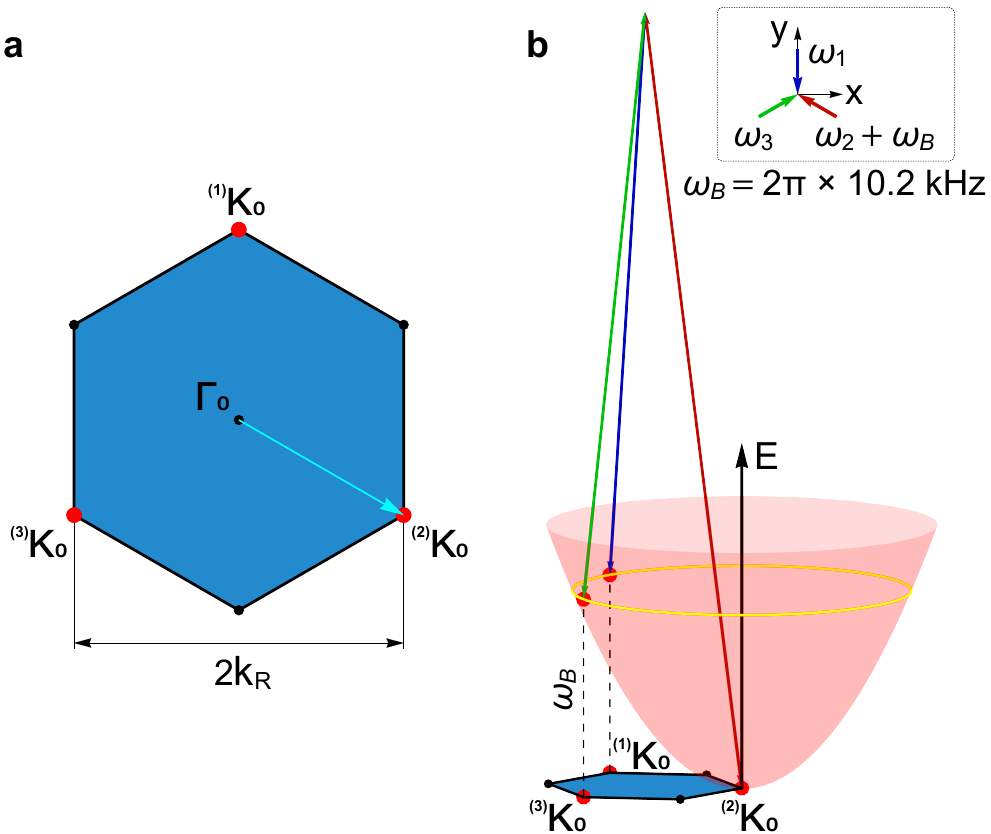}
\setcounter{figure}{3}
\renewcommand{\thefigure}{S\arabic{figure}}
\caption{\textbf{Schematic of the resonance coupling between the diffraction momenta at the three $^{(1)}K_0$, $^{(2)}K_0$, $^{(3)}K_0$ points.} $\bf{a}$~BEC is prepared at the Dirac point of the first Brillouin zone of $K_{0}$. $\bf{b}$~By adjusting the laser frequencies with $\Delta\omega_1=\Delta\omega_3=0$ and $\Delta\omega_2=4E_R/\hbar=2\pi\times 10.2$ kHz, the condensate is prepared at $K_0$ with quasi-momentum $\mathbf{q}_0=-\mathbf{k}_2$. The paraboloid represents the atomic kinetic energy. The atomic kinetic energy difference between two diffraction momenta is just the frequency difference between the two lasers for the two-photon process. The colors of three lattice beams correspond to that in Fig.~2a of the main text.}
\label{figS4}
\end{figure*}

By slightly adjusting the laser frequency, one can make the optical lattice moving with velocity $\mathbf{v}$~\cite{DStamper-Kurn}. In the lattice frame the atoms are transferred to some nonzero quasi-momentum $\mathbf{q}=-\frac{m}{\hbar}\mathbf{v}=-\frac{h}{4E_R}(\Delta\omega_1\mathbf{k}_1
+\Delta\omega_2\mathbf{k}_2+\Delta\omega_3\mathbf{k}_3)$, where $\mathbf{k}_1=-k_0\mathbf{e}_y$, $\mathbf{k}_2=k_0\big(-\frac{\sqrt{3}}{2}\mathbf{e}_x+\frac{1}{2}\mathbf{e}_y\big)$ and $\mathbf{k}_3=k_0\big(\frac{\sqrt{3}}{2}\mathbf{e}_x+\frac{1}{2}\mathbf{e}_y\big)$ are the three laser wave vectors for the honeycomb optical lattice, and $\Delta\omega_{1,2,3}$ are the frequency changes of the lasers in tens of kHz, which does not influence the wavelength and the lattice structure but changes the position of the atoms in the Brillouin zone.

Consider the case of moving the optical lattice along the direction of laser $\mathrm{k}_2$ with velocity $\mathbf{v}= \frac{\hbar\mathbf{k}_2}{m}$ by adjusting the laser frequencies with $\Delta\omega_1=\Delta\omega_3=0$ and $\Delta\omega_2=4E_R/h$. In this case, the condensate reaches the $K_0$ point of the first Brillouin zone boundary with quasi-momentum $\mathbf{q}_0=-\mathbf{k}_2$ in the lattice frame, as shown in Fig.~2b of the main text. With weak lattice depth, the momentum state $\mathbf{q}$ around $K_0$ is only resonantly coupled with the diffraction momenta $\mathbf{q}+\mathbf{G}_p$ with $p=(1,0)$ and $(1,1)$, which means the frequency difference between two lasers of a two-photon process is just the energy difference between the two diffraction momentum states, as shown in Fig.~S4. Defining $\mathbf{q}=\mathbf{q}_0+\mathbf{k}$, Eq.~(\ref{MomenHamiltonian}) is reduced to a three-band model with the Hamiltonian
\begin{equation}
\tilde{H}=\left(
\begin{array}{ccc}
\frac{\hbar ^{2}\left( \mathbf{k}-\mathbf{k}_{1}\right) ^{2}}{2m} & \Lambda
& \Lambda  \\
\Lambda  & \frac{\hbar ^{2}\left( \mathbf{k}-\mathbf{k}_{2}\right) ^{2}}{2m}
& \Lambda  \\
\Lambda  & \Lambda  & \frac{\hbar ^{2}\left( \mathbf{k}-\mathbf{k}_{3}\right)
^{2}}{2m}%
\end{array}%
\right),\label{H3x3}
\end{equation}
where the coupling strength $\Lambda=\frac{V_0}{9}$. With relatively weak lattice depth, this Hamiltonian can very well describe the $n=1,2,3$ Bloch bands in the vicinity of $K_0$ as well as the singular Dirac degeneracy between the $n=1$ and $n=2$ bands, as shown in Fig.~S5. This three-band model is equivalent to that of the cyclically Raman coupling scheme for spin-orbit coupling with near-resonant vector laser fields~\cite{GJuzelinas,JZhang}. Here the realization of cyclically Raman coupling with large-detuning scalar laser fields can effectively avoid the heating effect of single-photon resonance on the atoms. It should be noted that Eq.~(\ref{H3x3}) is only valid for a small momentum deviation $\mathbf{k}$ around $K_0$. This condition is satisfied due the small momentum spread of the condensate. In our simulation, we suppose the momentum distribution of the condensate around $K_0$ is described by a Gaussian function $\frac{1}{\rho\sqrt{2\pi}}e^{-\frac{(\mathbf{k}-\mathbf{q}_0)^2}{2\rho^2}}$, where $\rho$ denotes the momentum spread.

\begin{figure*}[tb]
\includegraphics[width=6in]{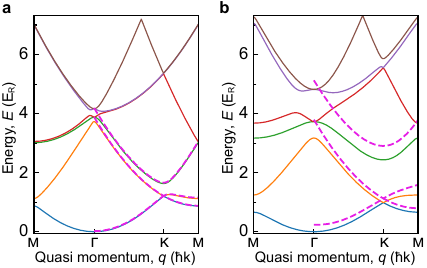}
\setcounter{figure}{4}
\renewcommand{\thefigure}{S\arabic{figure}}
\caption{\textbf{Three-mode approximation for the energy bands of the honeycomb optical lattice.} The bands with the lattice depth $V_0=2E_R$ and $V_0=4E_R$ are shown in $\mathbf{a}$ and $\mathbf{b}$, respectively. While the coloured curves show the bands calculated by the plane-wave expansion method, the magenta curves show the bands calculated from the effective three-band model.}
\label{figS5}
\end{figure*}

Making a unitary transformation
\begin{equation}
S=\frac{1}{\sqrt{3}}\left(
\begin{array}{ccc}
1 & 1 & 1 \\
e^{i\frac{2\pi }{3}} & e^{-i\frac{2\pi }{3}} & 1 \\
e^{-i\frac{2\pi }{3}} & e^{i\frac{2\pi }{3}} & 1%
\end{array}%
\right)
\end{equation}
one can transform the system from the bare-state representation with basis $\{|^{(1)}K_{0}\rangle,|^{(2)}K_{0}\rangle,|^{(3)}K_{0}\rangle\}$ to a dressed-state representation with basis $\{|\psi_1\rangle,|\psi_2\rangle,|\psi_3\rangle\}$. The effective Hamiltonian in the dressed-state representation then is given by
\begin{equation}
\bar{H}=\frac{\hbar ^{2}\mathbf{k}^{2}}{2m}+\frac{\hbar ^{2}k_{0}%
}{2m}\left(
\begin{array}{ccc}
0 & k_{y}+ik_{x} & k_{y}-ik_{x} \\
k_{y}-ik_{x} & 0 & k_{y}+ik_{x} \\
k_{y}+ik_{x} & k_{y}-ik_{x} & 0%
\end{array}%
\right) +\frac{\hbar^2k_{0}^2}{2m}+\left(
\begin{array}{ccc}
-\Lambda  & 0 & 0 \\
0 & -\Lambda  & 0 \\
0 & 0 & 2\Lambda
\end{array}%
\right).\label{SH3x3}
\end{equation}
Here the degenerate basis states $|\psi_1\rangle$ and $|\psi_2\rangle$ with the lowest energy $-\Lambda$ just correspond to the population on the sublattices $A$ and $B$ of the honeycomb lattice in the real space, respectively. The second term in the Hamiltonian $\bar{H}$ represents a extended version of the two-dimensional Rashba spin-orbit coupling in the SU(3) spin space, and can be rewritten as
\begin{equation}
\bar{H}_{\mathrm{so}}=\frac{\hbar ^{2}k_{0}}{2m}\mathbf{k}\cdot \mathbf{\Sigma},
\end{equation}
where the atom momentum $\mathbf{k}$ is coupled with the SU(3) pseudo-spin $\mathbf{\Sigma}=(\lambda_5-\lambda_2-\lambda_7)\mathbf{e}_x+(\lambda_1+\lambda_4+\lambda_6)\mathbf{e}_y$, in which $\lambda_s$ are the Gell-Mann matrices, i.e., the generators of the SU(3) group~\cite{GBArfken}
\begin{eqnarray}
\lambda _{1} &=&\left(
\begin{array}{ccc}
0 & 1 & 0 \\
1 & 0 & 0 \\
0 & 0 & 0%
\end{array}%
\right) ,\lambda _{2}=\left(
\begin{array}{ccc}
0 & -i & 0 \\
i & 0 & 0 \\
0 & 0 & 0%
\end{array}%
\right) ,\lambda _{3}=\left(
\begin{array}{ccc}
1 & 0 & 0 \\
0 & -1 & 0 \\
0 & 0 & 0%
\end{array}%
\right) ,  \nonumber \\
\lambda _{4} &=&\left(
\begin{array}{ccc}
0 & 0 & 1 \\
0 & 0 & 0 \\
1 & 0 & 0%
\end{array}%
\right) ,\lambda _{5}=\left(
\begin{array}{ccc}
0 & 0 & -i \\
0 & 0 & 0 \\
i & 0 & 0%
\end{array}%
\right) ,  \nonumber \\
\lambda _{6} &=&\left(
\begin{array}{ccc}
0 & 0 & 0 \\
0 & 0 & 1 \\
0 & 1 & 0%
\end{array}%
\right) ,\lambda _{7}=\left(
\begin{array}{ccc}
0 & 0 & 0 \\
0 & 0 & -i \\
0 & i & 0%
\end{array}%
\right) ,\lambda _{8}=\frac{1}{\sqrt{3}}\left(
\begin{array}{ccc}
1 & 0 & 0 \\
0 & 1 & 0 \\
0 & 0 & -2%
\end{array}%
\right) .  \label{GellMann}
\end{eqnarray}%
This type of SU(3) spin-orbit coupling goes beyond the scope of the traditional SU(2) spin-orbit coupling in condensed matter physics, and has been previously theoretically predicted in ultracold atoms by stimulated Raman coupling~\cite{VGalitski}. Here we first realize this model in a realistic experiment.

When the coupling strength $\Lambda$ is relatively stronger, one can take the spin-orbit coupling term of Eq.~(\ref{SH3x3}) as a perturbation around the zero momentum. As a result, the SU(3) spin-orbit coupling can be approximate to an SU(2) spin-orbit coupling up to second order, and the effective Hamiltonian becomes
\begin{equation}
\bar{H}\simeq\bar{H}^{(0)}+\bar{H}^{(1)}+\bar{H}^{(2)},\label{2x2Hamiltonian}
\end{equation}
where
\begin{eqnarray}
\bar{H}^{\left( 0\right) } &=&\frac{\hbar ^{2}\mathbf{k}^{2}}{2m}+\frac{\hbar^2k_{0}^2}{2m}-\Lambda,   \label{H0} \\
\bar{H}^{\left( 1\right) } &=&\frac{\hbar ^{2}k_{0}}{2m}\left(
\begin{array}{cc}
0 & \allowbreak k_{y}+ik_{x} \\
k_{y}-ik_{x} & 0%
\end{array}%
\right),   \label{H1} \\
\bar{H}^{\left( 2\right) } &=&-\frac{2E_R}{9\Lambda }\frac{\hbar ^{2}\mathbf{k}^{2}}{2m%
}-\frac{E_R}{9\Lambda }\frac{\hbar^2}{m}\left(
\begin{array}{cc}
0 & (k_{y}-ik_{x})^2 \\
(k_{y}+ik_{x})^2 & 0%
\end{array}%
\right).   \label{H2}
\end{eqnarray}
The first order Hamiltonian $H^{(1)}$ is just the familiar Rashba spin-orbit coupling and can be considered as an SU(2) spin lying in a two-dimensional circular pseudo-magnetic field
$\mathbf{B}\left( \mathbf{k}\right) =\frac{\hbar ^{2}k_{0}}{2m}k\left(\sin{\theta}\mathbf{e}_{x}- \cos{\theta}\mathbf{e}_{y}\right)$ with $\theta$ being the azimuthal angle, which wraps once around the degenerate point at $\mathbf{k}=0$, and the field strength is directly proportion to the distance away from the center. The second order Hamiltonian $H^{(2)}$ introduces a second-order spin-orbit coupling and can be considered as an SU(2) spin lying in a hexapole pseudo-magnetic field $\mathbf{B}\left( \mathbf{k}\right) =\frac{E_R}{9\Lambda}\frac{\hbar ^{2}k^2}{m}\left(\cos{2\theta}\mathbf{e}_{x}-\sin{2\theta}\mathbf{e}_{y}\right)$ in the momentum space, which wraps twice around the degenerate point~\cite{WKetterle}.

In addition, if we adjusting the laser frequencies with $\Delta\omega_1=\Delta\omega_3=0$ and $\Delta\omega_2=-4E_R/h$, the condensate will be transported to the $K'_0$ point of the Brillouin zone. Then the effective three-band model is written as
\begin{equation}
\tilde{H}=\left(
\begin{array}{ccc}
\frac{\hbar ^{2}\left( \mathbf{k}+\mathbf{k}_{1}\right) ^{2}}{2m} & \Lambda
& \Lambda  \\
\Lambda  & \frac{\hbar ^{2}\left( \mathbf{k}+\mathbf{k}_{2}\right) ^{2}}{2m}
& \Lambda  \\
\Lambda  & \Lambda  & \frac{\hbar ^{2}\left( \mathbf{k}+\mathbf{k}_{3}\right)
^{2}}{2m}%
\end{array}%
\right),\label{H3x3b}
\end{equation}
and by making a unitary transformation
\begin{equation}
S=\frac{1}{\sqrt{3}}\left(
\begin{array}{ccc}
1 & 1 & 1 \\
e^{-i\frac{2\pi }{3}} & e^{i\frac{2\pi }{3}} & 1 \\
e^{i\frac{2\pi }{3}} & e^{-i\frac{2\pi }{3}} & 1%
\end{array}%
\right),
\end{equation}
the Hamiltonian in the dressed-state representation is given by
\begin{equation}
\bar{H}=\frac{\hbar ^{2}\mathbf{k}^{2}}{2m}+\frac{\hbar ^{2}k_{0}%
}{2m}\left(
\begin{array}{ccc}
0 & -k_{y}+ik_{x} & -k_{y}-ik_{x} \\
-k_{y}-ik_{x} & 0 & -k_{y}+ik_{x} \\
-k_{y}+ik_{x} & -k_{y}-ik_{x} & 0%
\end{array}%
\right) +\frac{\hbar^2k_{0}^2}{2m}+\left(
\begin{array}{ccc}
-\Lambda  & 0 & 0 \\
0 & -\Lambda  & 0 \\
0 & 0 & 2\Lambda
\end{array}%
\right).\label{SH3x3b}
\end{equation}
Here the modified transformation matrix $S$ conserves the correspondence between the basis states $\psi_1$ and $\psi_2$ and the sublattices $A$ and $B$ as that of $K_0$. For the first-order pseudospin-1/2 approximation, we have
\begin{equation}
\bar{H}^{\left( 1\right) }=\frac{\hbar ^{2}k_{0}}{2m}\left(
\begin{array}{cc}
0 & -k_{y}+ik_{x} \\
-k_{y}-ik_{x} & 0%
\end{array}%
\right)
\end{equation}
which represents a Dresselhaus spin-orbit coupling, and can be considered as an SU(2) spin lying in a hyperbolic pseudo-magnetic field $\mathbf{B}\left( \mathbf{k}\right) =-\frac{\hbar ^{2}k_{0}}{2m}k\left(\sin{\theta}\mathbf{e}_{x}+\cos{\theta}\mathbf{e}_{y}\right)$ with an opposite chirality as that of $K_0$~\cite{AKGeim}.

\noindent \textbf{2. Generate vortex by gauge potential with the help of the harmonic trap}

Our experimental scheme can be considered as a momentum-space counterpart of topological manipulation of BECs using spatially varying magnetic fields~\cite{YShin,DSHall,DSHall2}. For a condensate with internal spin degrees of freedom in a spatially varying magnetic field, such as a quadrupole field, the single-particle Hamiltonian in the $\hat{z}$-quantized axis representation is written as
\begin{equation}
H_{r}=\frac{\mathbf{p}^2}{2m}+\mu _{B}g_{F}\mathbf{F}%
\cdot \mathbf{B}\left( \mathbf{r}\right)+\frac{1}{2}m\omega^2\mathbf{r}^2,
\end{equation}
where $\mathbf{F}=(F_{x}, F_{y}, F_{z})$ is the spin operator. A local unitary transformation $U(\mathbf{r})$ that rotates the quantization axis for the spin from the $\hat{z}$ direction to the local magnetic field direction $\hat{b}$ induces an effective gauge potential for the condensate due to its non-commutativity with the kinetic-energy operator. Then the transformed Hamiltonian in the $\hat{b}$-quantized axis representation is given by
\begin{equation}
H_{r}'=\frac{(\mathbf{p}+\mathbf{A})^2}{2m}+\mu _{B}g_{F}F_z|\mathbf{B}\left( \mathbf{r}\right)|+\frac{1}{2}m\omega^2\mathbf{r}^2.
\end{equation}
When the condensate is adiabatic prepared in one of the spin eigenstates, the local unitary transformation requires the atoms to feel a Berry phase along the closed path around the center of the magnetic field in the $\hat{z}$-quantized axis representation, which usually imprints real-space nontrivial topological objects, such as vortex, skyrmion and monople in the condensate~\cite{YShin,DSHall,DSHall2}.

In our experiment, the three-wave resonance of the atoms in the moving scalar honeycomb optical lattices generates a Rashba spin-orbit coupling between the two degenerate dressed spin states, which is equivalent to a momentum-space-varying circular magnetic field around the Dirac point as discussed above. Different from the case in real space, while this momentum-space pseudo-magnetic field commutes with the kinetic-energy operator, it does not commute with
the harmonic trap. Thus the harmonic trap can play the role similar to that of the kinetic-energy operator for generating gauge potential in real space, and is indispensable for the generation of momentum-space gauge potential.

In the spin-$1/2$ dressed-state representation for $K_0$, the effective Hamiltonian is given by
\begin{equation}
H_{p}=\frac{1}{2}m\omega^2\mathbf{r}^2+\mathbf{\sigma}\cdot \mathbf{B}(\mathbf{k})+\frac{\mathbf{p}^{2}}{2m},
\end{equation}
where the terms of uniform potentials are ignored. While the real-space harmonic potential can be written as a Laplace operator in the momentum space, the kinetic energy term can be considered as a momentum-space harmonic potential. Next we make a local unitary transformation $U(\mathbf{k})$, which rotates the spin quantization axis from the $z$-direction to the direction of the local pseudo-magnetic field, i.e., transforms the eigenstates of $\mathbf{\sigma}\cdot\hat{b}$ to the eigenstates of $\sigma_z$. This transformation is just a spin rotation around the axis $\hat{n}=(\hat{b}\times \hat{z})/|\hat{b}\times \hat{z}|=\left(-\cos\theta,-\sin\theta,0\right)$ by $\pi/2$, and the unitary transformation matrix for the spin-$\frac{1}{2}$ order parameter is given by
\begin{equation}
U=e^{i\frac{\pi }{2}\hat{n}\cdot \mathbf{\sigma}}=\frac{1}{\sqrt{2}}\left(
\begin{array}{cc}
1 & -ie^{-i\theta} \\
-ie^{i\theta} & 1%
\end{array}%
\right)
\end{equation}
The transformed Hamiltonian then is given by
\begin{equation}
H_{p}^{\prime }=U^{\dag }H_{p}U=\frac{1}{2}m\omega ^{2}\left(\mathbf{r}+\mathbf{A}\right) ^{2}+\sigma _{z}|\mathbf{B}(\mathbf{k})|+\frac{\mathbf{p}^{2}}{2m}.
\end{equation}
One finds that in the presence of the real-space harmonic trap this transformation introduces a gauge potential
\begin{equation}
\mathbf{A}=-iU^{\dag }\nabla _{\mathbf{k}}U=\mathbf{A}_{x}\sigma _{x}+\mathbf{A}%
_{y}\sigma _{y}+\mathbf{A}_{z}\sigma _{z}
\end{equation}
where the components of $\mathbf{A}$ are
\begin{eqnarray*}
\mathbf{A}_{x} &=&\frac{\sin \theta}{k}e_{\theta} \\
\mathbf{A}_{y} &=&-\frac{\cos \theta}{k}e_{\theta} \\
\mathbf{A}_{z} &=&\frac{1}{k}e_{\theta}
\end{eqnarray*}
In the experiment, by slowly adjusting the laser frequency from $\Delta\omega_2=0$ to $\Delta\omega_2=4E_R/h$, the atoms are adiabatically moved to the magnetic-field center. In this process, the atoms maintain in the same spin eigenstate with respect to the local $\hat{b}$-quantized axis. This means that there is no spin flips in this process, and the off-diagonal terms in $\bar{H}'$ can be ignored. As a result, the effective Hamiltonian in the adiabatic approximation is given by
\begin{equation}
\bar{H}_{\text{ad}}^{\prime }=\frac{1}{2}m\omega ^{2}\left( -i\nabla _{%
\mathbf{k}}+\frac{1}{k}e_{\theta }\sigma _{z}\right) ^{2}+\frac{1}{8k^{2}}%
m\omega ^{2}+\frac{\hbar ^{2}\mathbf{k}^{2}}{2m}+\sigma _{z}|\mathbf{B}|
\end{equation}
where the second term is the scalar potential resulting from $(\mathbf{A}_{x}\sigma _{x}+\mathbf{A}_{y}\sigma _{y})^2$.
The dynamics of the scalar wavefunction $\psi(\mathbf{k})$ for one spin state can be effectively simulated by the Hamiltonian
\begin{equation}
\bar{H}_{\mathrm{sc}}'=\frac{1}{2}m\omega ^{2}\left( -i\nabla _{\mathbf{k}}+\frac{1}{2k}e_{\theta
}\right) ^{2}+\frac{1}{8k^{2}}m\omega ^{2}+\frac{\hbar ^{2}\mathbf{k}^{2}}{2m%
}+\frac{\hbar ^{2}k_{0}}{2m}k.\label{Had}
\end{equation}
From Eq.~(\ref{Had}), the atom experiences an angular gauge potential
\begin{equation}
\mathbf{A}_{\mathrm{ad}}=\frac{1}{2k}e_{\theta}.
\end{equation}
Obviously, one finds that the corresponding gauge magnetic field is
\begin{equation}
\mathbf{B}=\mathbf{\nabla}\times\mathbf{A}_{\mathrm{ad}}=0,
\end{equation}
except the singular point at $\mathbf{k}=0$. However, the integral of the gauge potential $\mathbf{A}_{\mathrm{ad}}$ around a closed loop will lead to an effective magnetic flux
\begin{equation}
\Upsilon=\oint \hbar\mathbf{A}_{\mathrm{ad}}\cdot d\mathbf{l}=\pi\hbar,
\end{equation}
which accumulates an Aharonov-Bohm geometric phase factor $\exp{(i\pi)}$ in the wave function.

When BEC is accelerated from $\Gamma_{0}$ to $K_{0}$, the eigenstate $\Psi'_{s}=(\psi_{0}(\mathbf{k}),0)^{T}$ in a gauge transformed representation is prepared adiabatically. Here, $\psi_{0}(\mathbf{k})$ is the spin-independent scalar wave function in momentum space. The corresponding wavefunction in the dressed-state $z$-quantized representation is given by
\begin{equation}
\Psi_{s} =U\Psi'_{s} =\frac{1}{\sqrt{2}}\left(
\begin{array}{c}
1\\
-ie^{i\theta}%
\end{array}%
\right) \psi_{0}(\mathbf{k}),
\end{equation}
the wavefunction in the bare-state representation is written as
\begin{equation}
\Phi =S\Psi =\frac{1}{\sqrt{6}}\left(
\begin{array}{c}
1-ie^{i\theta } \\
e^{i\frac{2\pi }{3}}-ie^{-i\frac{2\pi }{3}}e^{i\theta } \\
e^{-i\frac{2\pi }{3}}-ie^{i\frac{2\pi }{3}}e^{i\theta }%
\end{array}%
\right) \psi_{0}(\mathbf{k})=\frac{1}{\sqrt{6}}\left(
\begin{array}{c}
e^{i\beta_A^{1}}+e^{i\beta_B^{1}}e^{i\theta } \\
e^{i\beta_A^{2}}+e^{i\beta_B^{2}}e^{i\theta } \\
e^{i\beta_A^{3}}+e^{i\beta_B^{3}}e^{i\theta }%
\end{array}%
\right) \psi_{0}(\mathbf{k}),\label{Psibare}
\end{equation}
where $\beta_A^{a}$ and $\beta_B^{a}$ ($a=1,2,3$) are constant phases. The spatially varied phase factor $e^{i\theta}$ is responsible for the vortex formation as shown in Figs.~3b and 3c of the main text.

\begin{figure*}[tb]
\includegraphics[width=4in]{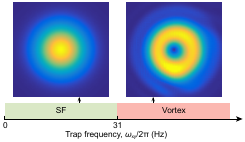}
\setcounter{figure}{5}
\renewcommand{\thefigure}{S\arabic{figure}}
\caption{\textbf{Theoretical prediction of the superfluid and vortex phase.} Atomic density of the spin-independent scalar wave function $\psi_{0}(\mathbf{k})$ as a function of the harmonic trapping frequency after the adiabatic evolution. A strong enough harmonic trapping frequency can induce a density hole in the atoms, which is critical for the generation of vortex excitation. Here we suppose the lattice depth is moderate (e.g. $V_0=12 E_R$) to ensure that the pseudo-spin 1/2  model is effective. The time for the adiabatic evolution is fixed about 1 ms in our calculation.}
\label{figS6}
\end{figure*}

\begin{figure*}[tb]
\includegraphics[width=4.5in]{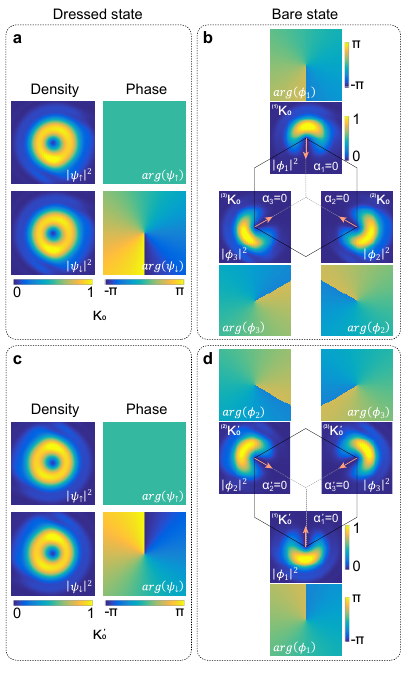}
\setcounter{figure}{6}
\renewcommand{\thefigure}{S\arabic{figure}}
\caption{\textbf{The amplitude and phase of the wavefunction for the vortex states at $K_0$ and $K'_0$.} The vortex states for $K_0$ in the dressed-state and bare-state representations are shown in $\mathbf{a}$ and $\mathbf{b}$, respectively. The vortex states for $K'_0$ in the dressed-state and bare-state representations are shown in $\mathbf{c}$ and $\mathbf{d}$, respectively. The harmonic trapping frequency in the $x$-$y$ direction is $2\pi\times40$~Hz, and the depth of optical lattice is 12~$E_{R}$.}
\label{figS7}
\end{figure*}

The spin-independent scalar wave function $\psi_{0}(\mathbf{k})$ can be calculated by adiabatically moving a Gaussian wave package from $\left(-\frac{\sqrt{3}}{2}k_0,\frac{1}{2}k_0\right)$ to the magnetic-field center $\left(0,0\right)$ according to the time-dependent Schr\"{o}dinger equation
\begin{equation}
i\frac{\partial\psi_{0}(\mathbf{k},t)}{\partial t}=\bar{H}_{\mathrm{sc}}'\psi_{0}(\mathbf{k},t).
\end{equation}
It is found that, beside the gauge potential $\mathbf{A}$ induced by the harmonic trap for the atoms, the adiabatic evolution with a strong enough harmonic trapping frequency can produce a density hole in the wave package, as shown in the right panel of Fig.~S6, which is is critical for the generation of vortex excitation. When the trapping frequency is weak, the density hole is hard to be produced, and the density of the atoms still behave as a Gaussian wave package after the adiabatic evolution, as shown in the left panel of Fig.~S6. In this case, the phase singularity of the vortex can not enter into the atoms due to enormous energy cost. When the trapping frequency is strong, the adiabatic evolution produces a density hole in the center of the atoms. As a result, the the phase singularity can easily enter into the atoms with less energy cost.

From Fig.~3c and Fig.~S7b, we find that the vortex state for the $K_0$ point in the bare-state representation are not rotational symmetric, but there is a breach in the density ring with the opening direction being towards the Brillouin zone centre with $\alpha_{a}=0$ ($a=1,2,3$). The vortex state prepared adiabatically at the Dirac point the $K_0$ point is expressed as $\Psi_{s}=(1,- i e^{i\theta})^{T}$ in the pseudo-spin 1/2 representation and the projection to three momentum components $^{(1)}K_{0}$, $^{(2)}K_{0}$ and $^{(3)}K_{0}$ in TOF image is connected to the quasi-spin components of this state as $\phi_{1}=\frac{1}{\sqrt{3}}\left( \psi _{\uparrow}+\psi _{\downarrow}\right)$, $\phi_{2}=\frac{1}{\sqrt{3}}\left( e^{i\frac{2\pi }{3}}\psi _{\uparrow}+e^{-i\frac{2\pi}{3}}\psi _{\downarrow}\right)$, and $\phi_{3}=\frac{1}{\sqrt{3}}\left( e^{-i\frac{2\pi}{3}}\psi _{\uparrow}+e^{i\frac{2\pi }{3}}\psi _{\downarrow}\right)$. Therefore, the zero-density angle $\alpha_{a}$ can be calculated as $\alpha_a=\beta_{A}^{a}-\beta_{B}^{a}+\pi-\alpha_{a0}=0$. Here $\alpha_{a0}$ (with $\alpha_{10}=\frac{3\pi}{2}$, $\alpha_{20}=\frac{3\pi}{2}+\frac{4\pi}{3}$ and $\alpha_{30}=\frac{3\pi}{2}-\frac{4\pi}{3}$) are the angular positions of the three Dirac points $^{(1)}K_{0}$, $^{(2)}K_{0}$ and $^{(3)}K_{0}$. This results from the specific $C_3$ symmetry of the effective Hamiltonian in Eq.~(\ref{H3x3}).

When the condensate is adiabatically transported at the $K'_0$ point, the wavefunction in the dressed-state $z$-quantized representation is given by
\begin{equation}
\Psi_{s} =\frac{1}{\sqrt{2}}\left(
\begin{array}{c}
1\\
-ie^{-i\theta}%
\end{array}%
\right) \psi_{0}(\mathbf{k}),
\end{equation}
and the wavefunction in the bare-state representation is written as
\begin{equation}
\Phi =\frac{1}{\sqrt{6}}\left(
\begin{array}{c}
1-ie^{-i\theta } \\
e^{-i\frac{2\pi }{3}}-ie^{i\frac{2\pi }{3}}e^{-i\theta } \\
e^{i\frac{2\pi }{3}}-ie^{-i\frac{2\pi }{3}}e^{-i\theta }%
\end{array}%
\right) \psi_{0}(\mathbf{k})=\frac{1}{\sqrt{6}}\left(
\begin{array}{c}
e^{i\beta_A^{1}}+e^{i\beta_B^{1}}e^{-i\theta } \\
e^{i\beta_A^{2}}+e^{i\beta_B^{2}}e^{-i\theta } \\
e^{i\beta_A^{3}}+e^{i\beta_B^{3}}e^{-i\theta }%
\end{array}%
\right) \psi_{0}(\mathbf{k}).\label{Psibare2}
\end{equation}
Here the zero-density angle $\alpha'_{a}$ can be calculated as $\alpha'_a=\beta_B^{a}-\beta_A^{a}+\pi-\alpha'_{a0}=0$, where $\alpha'_{a0}$ (with $\alpha'_{10}=\frac{\pi}{2}$, $\alpha'_{20}=\frac{\pi}{2}+\frac{4\pi}{3}$ and $\alpha'_{30}=\frac{\pi}{2}-\frac{4\pi}{3}$) are the angular positions of the three Dirac points $^{(1)}K'_{0}$, $^{(2)}K'_{0}$ and $^{(3)}K'_{0}$. In this case, we find that the opening direction of the density of the vortex states is also towards the Brillouin zone centre, but the circulating direction of the vortex phase is opposite, as shown in Fig.~S7d.

\noindent \textbf{3. Generate vortex by coherent spin flip}

\begin{figure*}[htb]
\includegraphics[width=6in]{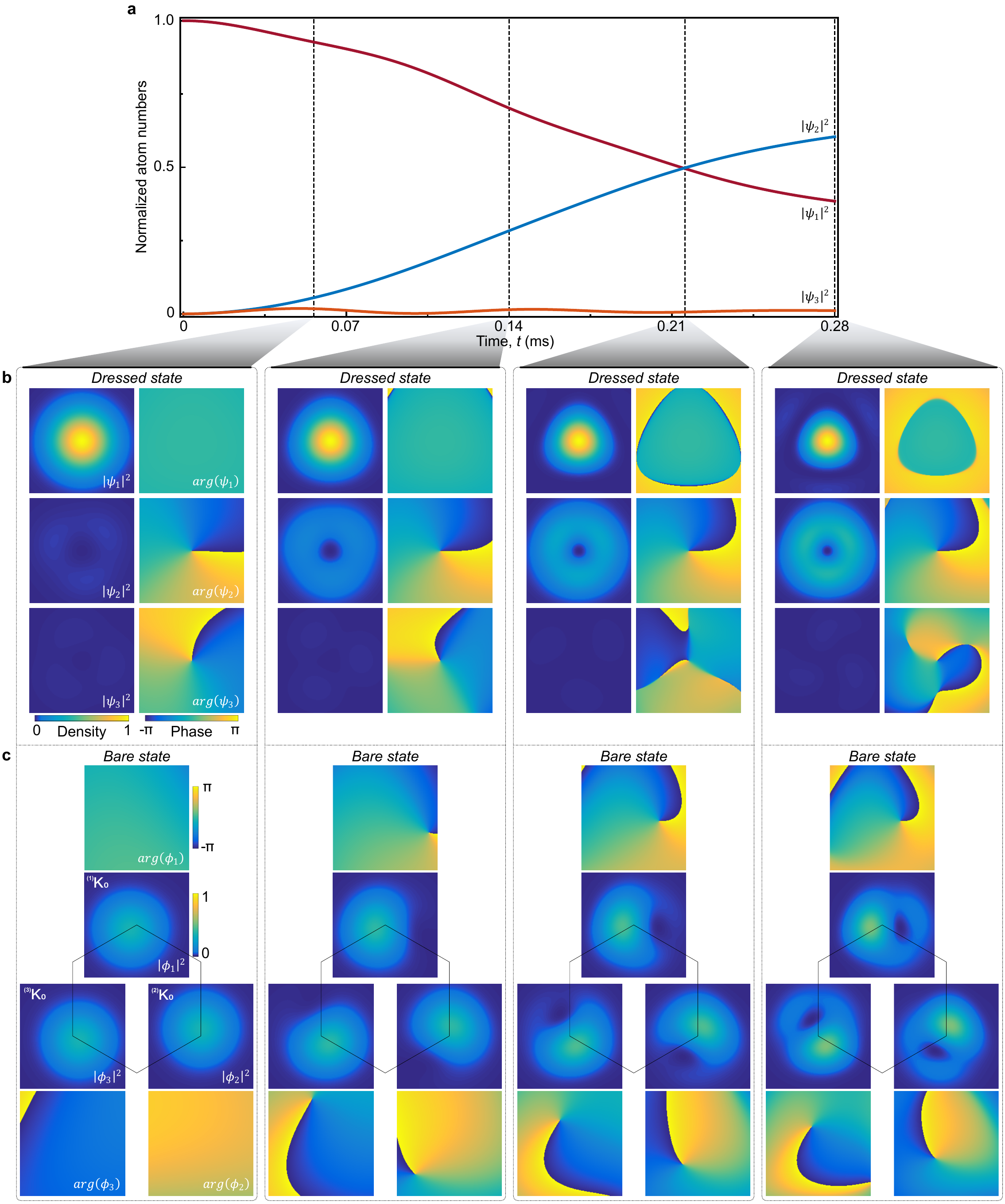}
\setcounter{figure}{7}
\renewcommand{\thefigure}{S\arabic{figure}}
\caption{\textbf{Dynamic generation of quantized vortex under the spin-orbit coupling at $K_0$.} $\mathbf{a}$ The population evolution of the atoms between the three spin components in the dressed-state representation. $\mathbf{b}$ and $\mathbf{c}$ show the time evolution of the density and phase distributions of the spin components in the dressed- and bare-state representation respectively. The depth of optical lattice is 12~$E_{R}$ in the calculation.}
\label{figS8}
\end{figure*}

In this section, by numerical simulation we show the conventional method to create vortices near the Dirac point based on dynamical process of coherent spin flip. The spinor order parameter in the dressed-state representation of Eq.~(\ref{SH3x3}) is given by $(\psi_1,\psi_2,\psi_3)^{T}$. Consider the condensate is initially prepared in one of the spin component $\psi_1$ around the Dirac point, and we suppose the momentum distribution of the condensate is described by a Gaussian function $\frac{1}{\rho\sqrt{2\pi}}e^{-\frac{(\mathbf{k}-\mathbf{q}_0)^2}{2\rho^2}}$, where $\rho$ denotes the momentum spread. The dynamical evolution of the order parameters according to the Hamiltonian of Eq.~(\ref{SH3x3}) is shown in Fig.~S8. In the case of relatively strong coupling strength $\Lambda$, the energy level of the component $\psi_3$ is far away from $\psi_1$ and $\psi_2$, thus most atoms are transferred to the component of $\psi_2$ as shown in Fig.~S8a.

\begin{figure*}[htb]
\includegraphics[width=6in]{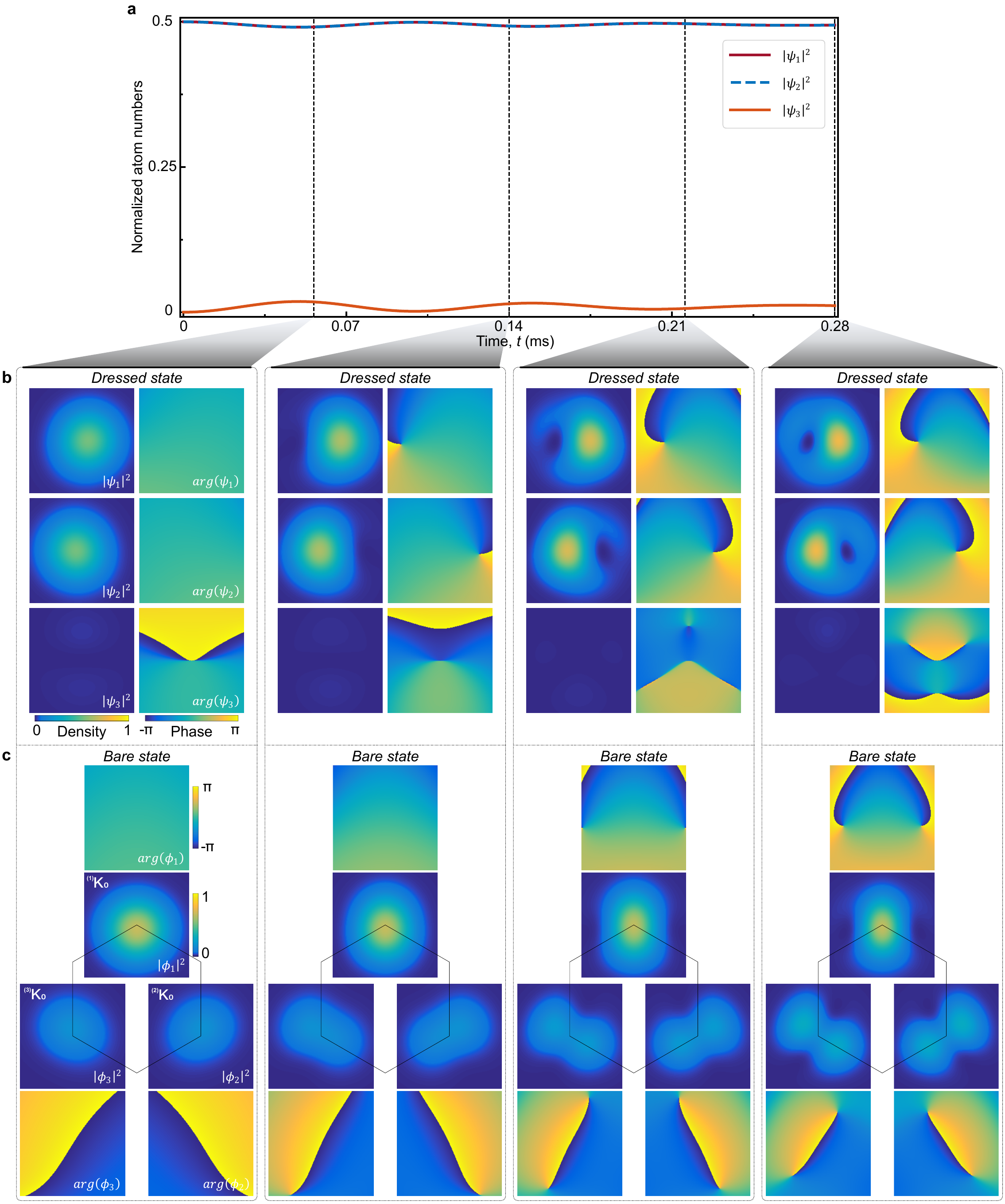}
\setcounter{figure}{8}
\renewcommand{\thefigure}{S\arabic{figure}}
\caption{\textbf{Dynamical evolution of the condensate initially prepared in an equal-weight superposition of the two degenerate spin components at the Dirac point of $K_0$.} $\mathbf{a}$ The population evolution of the atoms between the three spin components in the dressed-state representation. $\mathbf{b}$ and $\mathbf{c}$ show the time evolution of the density and phase distributions of the spin components in the dressed- and bare-state representation respectively. The depth of optical lattice is 12~$E_{R}$ in the calculation.}
\label{figS9}
\end{figure*}

Under spin-orbit coupling, the transfer of the atoms between the pseudo-spin components turns the spin angular momentum into orbital angular momentum, which leads to a quantized vortex in the transferred component, as shown in Fig.~S8b. It is also found that the density distribution of the vortex state has $\mathrm{C}_3$ symmetry, which results from the SU(3) spin-orbit coupling, and is in contrast to the vortex with $\mathrm{SO}(2)$ rotational symmetry induced by an SU(2) spin-orbit coupling~\cite{CWu}. The vortex dynamics in the bare-state representation is more different. While the vortex is directly formed in the center of the condensate in the dressed-state representation, the vortices gradually enter the condensate from the surface in the bare-state representation as shown in Fig.~S8c.

We find that the dynamic generation of vortices significantly depends on the initial state of the evolution, and it can not be guaranteed that an arbitrary initial state can generate evident vortex structures in the experimentally observable bare-state representation. For example, in most case of the realistic experiment, the quenching of the lattice velocity always transports the atoms to the Dirac point with a superposition of the two degenerate dressed spin states. In this case, while the subsequent dynamical evolution generally can generate evident vortices with density hole in the dress-state representation, only hidden vortices emerges around the surface of the condensates without evident density hole to be observed in the bare-state representation, as shown in Fig. S9.

\end{widetext}

\end{document}